\documentclass[aps,prl,twocolumn,superscriptaddress,showpacs,nofootinbib,longbibliography]{revtex4-1}
\usepackage{graphicx}
\usepackage{dcolumn}
\usepackage{bm}
\usepackage{hyperref}
\usepackage{upgreek}
\usepackage{color}
\usepackage{soul}
\usepackage{epstopdf}
\usepackage{units}
\usepackage{longtable}
\usepackage{floatrow}
\usepackage{latexsym}
\usepackage{gensymb}
\usepackage{floatrow}
\usepackage{mhchem}

\begin{document}

\title{Balanced Quantum Hall Resistor}

\author{K. M. Fijalkowski}
\email[email:]{kajetan.fijalkowski@physik.uni-wuerzburg.de}
\author{N. Liu}
\author{M. Klement}
\author{S. Schreyeck}
\author{K. Brunner}
\author{C. Gould}
\email[email:]{charles.gould@physik.uni-wuerzburg.de}
\author{L. W. Molenkamp}
\affiliation{Faculty for Physics and Astronomy (EP3), Universit\"at W\"urzburg, Am Hubland, D-97074, W\"urzburg, Germany}
\affiliation{Institute for Topological Insulators, Am Hubland, D-97074, W\"urzburg, Germany}

\begin{abstract}
The quantum anomalous Hall effect in magnetic topological insulators has been recognized as a promising platform for applications in quantum metrology. The primary reason for this is the electronic conductance quantization at zero external magnetic field, which allows to combine it with the quantum standard of voltage.
Here we demonstrate a measurement scheme that increases the robustness of the zero magnetic field quantum anomalous Hall resistor, allowing for higher operational currents. This is achieved by simultaneous current injection into the two disconnected perimeters of a multi-terminal Corbino device to balance the electrochemical potential between the edges, screening the electric field that drives back-scattering through the bulk, and thus improving the stability of the quantization at increased currents. This approach is not only applicable to devices based on the quantum anomalous Hall effect, but more generally can also be applied to existing quantum resistance standards that rely on the integer quantum Hall effect.
\end{abstract}

\maketitle

\section[Introduction]{Introduction}
The past few decades witnessed the rise of topological insulators in condensed matter physics \cite{Kane2005,Hasan2010,konig2007}, with the quantum anomalous Hall effect (QAHE) \cite{Yu2010,Nomura2011,Chang2013} demonstrating the interesting consequences of the interplay between topology and magnetism. This effect has opened up interesting avenues of academic research into its underlying magnetic and electrodynamic properties \cite{Nomura2011,Lachman2015,Grauer2015,Liu2016,Yasuda2016,tokura1,Grauer2017,Fijalkowski2020,Fijalkowski2021a,Fijalkowski2022}. Technologically relevant is that the effect is characterized by a universal electrical conductance robustly quantized to $e^2/h$. The fact that this quantization persists even at zero external magnetic field, brings promise for potential applications in quantum resistance metrology \cite{Goetz2018,Fox2018,Okazaki2020,Okazaki2022,Rodenbach2023}.

An important long-term goal in quantum metrology is to combine the quantum voltage standard based on the AC Josephson effect \cite{Joesphson1962} with the quantum resistance standard based on the quantum Hall effect \cite{Klitzing1980}, into a single reference instrument. This would provide simultaneous access to the von-Klitzing and the Josephson constants ($R_{\mathrm{K}} = h/e^2$ and $K_{\mathrm{J}} = 2e/h$), from which the elementary charge $e$ and Planck's constant $h$ can be determined, thus providing a universal quantum electrical standard. This is especially important in the context of the recent redefinition of the kilogram in terms of Planck's constant $h$ \cite{BIPM}, and to related experiments aimed at tracing the kilogram, such as the elaboration of a Kibble balance \cite{Kibble1976,Stock2012}. 

When compared to the ordinary quantum Hall effect, the obvious advantage of the QAHE is that it eliminates the need for an external magnetic field. The two main obstacles to the QAHE entering mainstream metrological use are that observation of its perfect quantization is currently limited to too low temperature and too small measurement current. On the issue of temperature, metrology grade measurements of the QAHE have thus far been limited to dilution fridge temperatures \cite{Goetz2018,Fox2018,Okazaki2020,Okazaki2022}. Nevertheless, evidence that the edge states themselves survive to much higher temperatures (up to the Curie temperature of some few tens of Kelvin \cite{Chang2015b,Yasuda2020,Fijalkowski2021b}) provides hope that the liquid $^4$He temperatures commonly used for Josephson quantum voltage standards can be reached through material optimization.  

Here, we focus on the second obstacle; that of the measurement current. The issue here is that the cryogenic current comparator (CCC) systems \cite{Harvey1972} typically used for metrological measurements work optimally at currents of tens of microamps \cite{Ribeiropalau2015}, and when current is restricted to lower levels, the accuracy of the instrument is reduced \cite{Goetz2018}. It is therefore beneficial for metrologically relevant devices to be able to operate at currents approaching the tens of microamps regime. Thus far, metrological level quantization of the QAHE at zero external magnetic field has been limited to currents below some 100 nA, which limited the quantization precision to the 10$^{-6}$-10$^{-7}$ level \cite{Goetz2018,Fox2018,Okazaki2020}. An operation current of 1 $\mu$A has been achieved with the help of $\sim$200 mT field produced by a permanent magnet \cite{Okazaki2022}, which improved this number to the 10$^{-8}$ level, but such a field hampers integrability with the superconducting devices needed for determining the Josephson constant. The aforementioned limitations prevent the QAHE from reaching the precision level of 10$^{-11}$, routinely achieved using the integer quantum Hall effect \cite{Ribeiropalau2015}.

In this Article we present a measurement scheme that significantly improves the robustness of zero magnetic field QAHE at higher currents \cite{Patent}. This is achieved by simultaneous current injection into the two disconnected perimeters of a multi-terminal Corbino device. In this geometry, the electric field created between the edges of the sample is compensated, suppressing back-scattering through the bulk. It is well worth noting that the same measurement method can also be applied to ordinary integer quantum Hall devices.

\begin{figure*}
\includegraphics[width=4in]{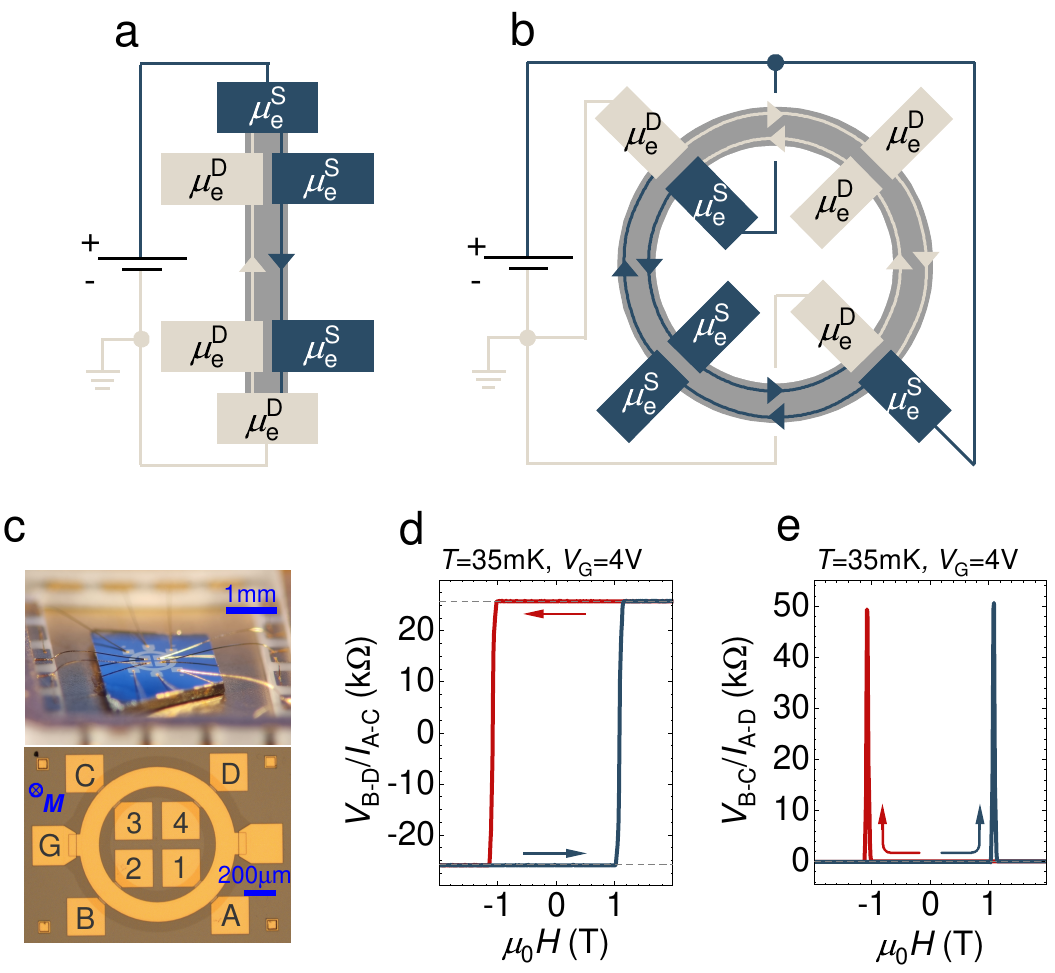}
\caption{
\textbf{Basic principle behind the electrochemical potential balancing method and characterization of the device.} a) Schematic of a Hall bar device under applied bias voltage. The dark blue line depicts the high electrochemical potential edge of the device ($\mu_{\mathrm{e}}^{\mathrm{S}}$), and the grey line shows the low electrochemical potential edge state ($\mu_{\mathrm{e}}^{\mathrm{D}}$). b) The same for an optimally double-sourced multi-terminal Corbino device. c) Top: Photograph of the bonded multi-terminal Corbino quantum anomalous Hall device; Bottom: Optical microscope image of the same device (prior to bonding). The letters A-D label contacts along the outer perimeter, numbers 1-4 those along the inner perimeter. G denotes the gate contact. d-e) Basic characterization of the device, without any balancing. d) Four-terminal Hall resistance ($V_{\mathrm{B-D}}$/$I_{\mathrm{A-C}}$) measured along the outer perimeter as a function of a perpendicular-to-plane external magnetic field. e) the corresponding longitudinal resistance ($V_{\mathrm{B-C}}$/$I_{\mathrm{A-D}}$) measurement. The colored arrows give the magnetic field sweep direction. Measurements in (d-e) are done at 21.21 Hz with a bias voltage of 100 $\mu$V, at a sample temperature of 35 mK, and with 4V applied to the gate. The horizontal dashed lines represent the resistance values expected for a perfect QAHE: $\pm$$h/e^2$ for the Hall resistance in (d), and 0 for the longitudinal resistance in (e). 
}
\label{fig:Fig1}
\end{figure*}

\begin{figure*}
\includegraphics[width=\columnwidth]{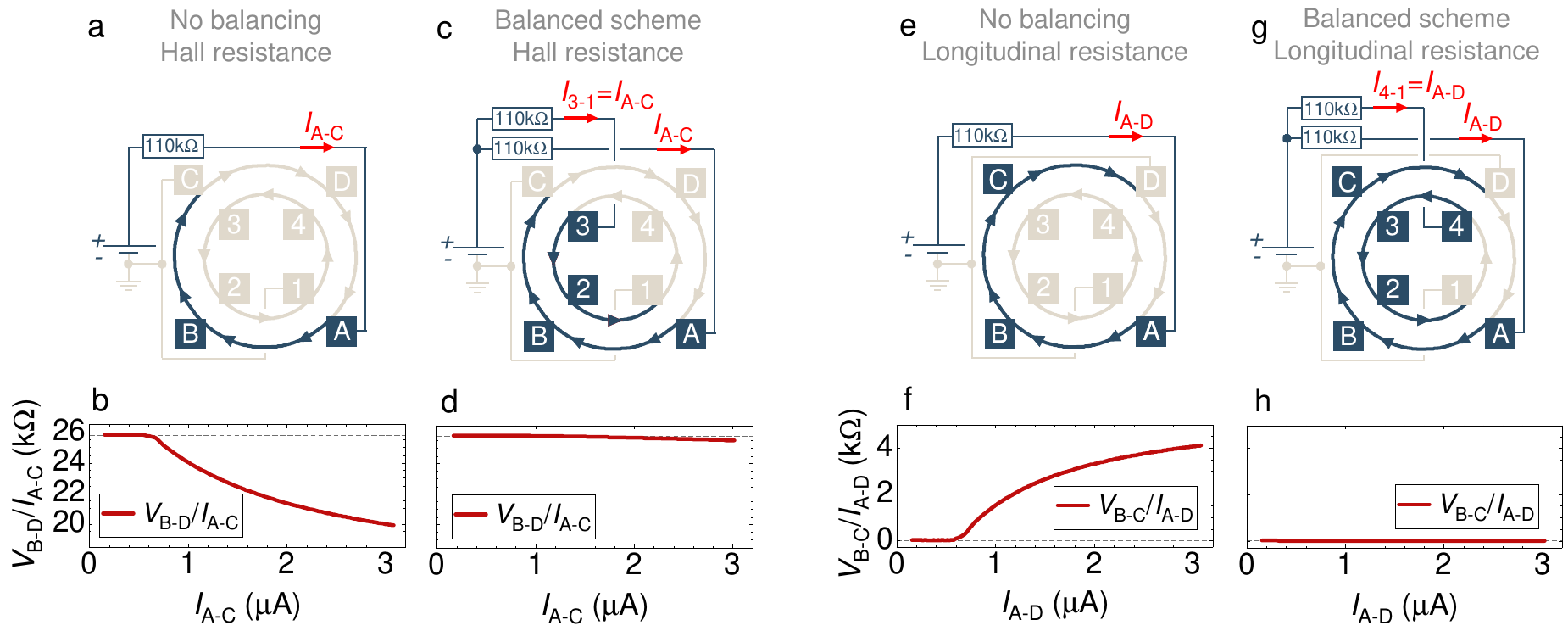}
\caption{
\textbf{Electrochemical potential balancing.} a) Circuit diagram for a reference measurement of the Hall resistance $V_{\mathrm{B-D}}$/$I_{\mathrm{A-C}}$ without electrochemical potential balancing. The outer perimeter is electrically biased, driving a DC current  $I_{\mathrm{A-C}}$ in the outer perimeter, while the inner perimeter is grounded via contact 1. The dark blue color indicates the high electrochemical potential part of the circuit, whereas light grey shows low potential. b) A measurement of $V_{\mathrm{B-D}}$/$I_{\mathrm{A-C}}$ for this configuration, as a function of current $I_{\mathrm{A-C}}$. c-d) Same for the Hall resistance $V_{\mathrm{B-D}}$/$I_{\mathrm{A-C}}$ measured with electrochemical potential balancing: An additional DC current $I_{\mathrm{3-1}}$ (of the same magnitude as $I_{\mathrm{A-C}}$) is passed at the inner perimeter, in order to balance the electrochemical potentials along the ring. e-h) Analogous, but for the longitudinal resistance $V_{\mathrm{B-C}}$/$I_{\mathrm{A-D}}$:  (e-f), without, and (g-h) with electrochemical potential balancing. The horizontal dashed lines in (b,d,f,h) show the  expected values for ideal quantum anomalous Hall edge states. All data is collected at zero external magnetic field, a temperature of 30 mK, and an applied gate voltage of 4V, to tune the Fermi level to observe the quantization plateau. 
}%
\label{fig:Fig2}%
\end{figure*}

\section[Inter-perimeter electrochemical potential balancing]{Inter-perimeter electrochemical potential balancing}

The basic idea of electrochemical potential balancing comes from realising that, in macroscopic devices (i.e. devices where the separation between edge states is larger than all other relevant length scales, such as the effective width of the edge state, the magnetic length, and the screening length), the breakdown mechanism for both the ordinary quantum Hall and quantum anomalous Hall edge state transport  is driven by the electric field between opposite edges of the device \cite{Komiyama1985,Komiyama2000,Kawamura2017,Fox2018,Rodenbach2021,Lippertz2022}.

The details of how this electric field is distributed between the edges \cite{Mccormick1999} can depend on the exact nature of the edge electrostatic potential reconstruction into compressible and incompressible regions and the associated screening \cite{Chklovskii1992}, and can be quite rich. The essence nevertheless remains that as the electric field between edge states at different potentials exceeds a critical value, electrons are driven by this field to traverse the bulk and back-scatter to the  opposite edges of the device, causing a breakdown in the quantization of the Hall signal. For the discussion of this paper and in the limit of large devices, the single-electron picture of purely one dimensional edge states with no special screening properties, captures the essence of the relevant mechanism, and is sufficient.  

Fig. 1a-b illustrates the basic advantage of our measurement scheme in this simple picture. In a Hall bar geometry in the edge state regime, the electrochemical potential along the edges of the device is distributed as presented schematically in Fig. 1a. A bias voltage is applied to the source contact,  whereas the drain contact is grounded. The edge state equilibrates the electric potential along one edge to that of the source contact, and the other edge to that of the drain. The edge state drawn in dark blue is at high electrochemical potential, whereas the one shown in light grey represents low (here grounded) electrochemical potential. As the measurement current is increased by applying a larger bias between source and drain, the electrochemical potential difference between the two edges of the sample increases. This increases the inter-edge back-scattering probability, and eventually leads to breakdown of the perfect conductance quantization. 

A multi-terminal Corbino geometry, first used in \cite{Fijalkowski2021b}, is pictured in Fig. 1c. In the limit of an insulating bulk, the inner and the outer perimeter can independently be used to measure the QAHE. On the outer (inner) perimeter, one can determine the Hall voltage by passing, for example, a current from A to C (1 to 3) and then measuring the resulting voltage between B and D (2 and 4). 

Making use of both perimeters allows the measuring scheme sketched in Fig. 1b. The source and (grounded) drain contacts along each perimeter are chosen in such a way that for one part of the circumference, the edge states running along both edges are at high electrochemical potential, while for the remainder of the circumference, both the edges are at the grounded electrochemical potential. This biasing method prevents the build up of an electric field between the inner and outer edge, and thus suppresses electric field induced breakdown mechanisms.

\section[Proof of concept]{Proof of concept}
The device is patterned using standard optical lithography methods, from a magnetic topological insulator V$_{0.1}$(Bi$_{0.2}$Sb$_{0.8}$)$_{1.9}$Te$_3$ layer, grown by molecular beam epitaxy (MBE) on a Si(111) substrate \cite{Winnerlein2017}. The outer diameter of the ring is 1 mm, the separation between the inner and outer perimeters is 100 $\mu$m, and the constriction defining each contact is 50 $\mu$m wide. The magnetic topological insulator layer has a thickness of 8.2 nm, and is capped with a 10 nm thick protective layer of amorphous Te, which is insulating. The sample is fitted with a ring shaped top gate, labeled G in Fig. 1c, which is comprised of a 15 nm thick atomic layer deposition (ALD) AlOx film and a 100 nm thick layer of Au.

The material is optimized to have a fully insulating bulk (when properly gated, and below 100 mK), and thus excellent anomalous Hall resistance quantization \cite{Grauer2015,Goetz2018}. The sample is mounted in a top-loader type dilution refrigerator with a modest cooling power of 400 $\mu$W at 100 mK, and base temperature of about 30 mK. A gate voltage of 4V is applied to tune the Fermi level and adjust the sample onto the QAHE plateau (see Figs. 1d and 1e where the Hall $V_{\mathrm{B-D}}$/$I_{\mathrm{A-C}}$ and longitudinal $V_{\mathrm{B-C}}$/$I_{\mathrm{A-D}}$ resistances are plotted as a function of the external out-of-plane magnetic field, showing a perfect QAHE). All other measurements in this paper are done in zero external magnetic field.

To test the effectiveness of the balancing method, first we perform a reference experiment without any balancing. Fig. 2a shows a circuit schematic for a measurement where the DC bias voltage is applied to the outer perimeter (between contacts A and C, which results in a DC current $I_{\mathrm{A-C}}$). The inner perimeter is grounded via contact 1. With both contacts C and 1 grounded, for the half of the circumference that includes contacts B/2, the electrochemical potential difference between the inner and outer edges is proportional to the bias applied at contact A, and thus proportional to the measurement current. This situation is equivalent to that of a measurement on a Hall bar. 

Fig. 2b shows a corresponding four terminal Hall resistance $V_{\mathrm{B-D}}$/$I_{\mathrm{A-C}}$ measurement along the outer perimeter, as a function of the current $I_{\mathrm{A-C}}$. At low currents this resistance is quantized to $h/e^2$. At a current $I_{\mathrm{A-C}}$ exceeding about 620 nA (with a relative deviation from quantization of about 1$\%$ observed at a current of about 680 nA), the critical electrochemical potential difference between the inner and outer edge is reached, and an abrupt breakdown of quantization is observed.

In Figs. 2c and 2d we turn to the balanced scheme. Fig. 2c shows a circuit diagram schematic, with an additional DC current $I_{\mathrm{3-1}}$ (of nominally the same magnitude as $I_{\mathrm{A-C}}$; both currents are measured to be equal to within about 1 $\%$) flowing into the inner perimeter at contacts 3 and 1, while C and 1 are grounded. A 110 k$\Omega$ resistor is placed in series with each perimeter to minimize the effects of contact resistance variations along the device (contact resistances are of the order of a few hundred ohms in this device). In this scenario, as the edge states equilibrate the electrochemical potentials, contacts 4 and D adjust to the electrochemical potential determined by the circuit ground, whereas contacts B and 2 are at the electrochemical potential determined by the source contacts, A and 3 respectively. As the two currents $I_{\mathrm{A-C}}$ and $I_{\mathrm{3-1}}$ are equal in magnitude, the electrochemical potential at contacts A and 3 is equal, leading to an electrochemical potential balance between the inner and outer edge at all points around the circumference of the ring. This is of course independent of the amplitude of the measurement current. 

The Hall resistance $V_{\mathrm{B-D}}$/$I_{\mathrm{A-C}}$ measured in the configuration of Fig. 2c is plotted in Fig. 2d as a function of the measurement current $I_{\mathrm{A-C}}$. It is clear that the electric field induced breakdown observed in Fig. 2b is effectively suppressed, allowing for larger current to be injected into the sample before a deviation from quantization is observed (a relative deviation from quantization of 1$\%$ is observed at a current of about 2.85 $\mu$A). Importantly, the nature of the departure from quantization in Fig. 2d is very different from that of Fig. 2b. Whereas in Fig. 2b, an abrupt departure is observed, characteristic of the critical electrochemical potential difference having been reached, in Fig. 2d, the Hall resistance moves away from quantization smoothly and gradually and at higher current values. We will show below that the breakdown of perfect quantization in this case can be attributed to thermal activation of bulk conductance caused by Joule heating.    

In Figs. 2e-h we turn to an equivalent analysis for the four-terminal longitudinal resistance $V_{\mathrm{B-C}}$/$I_{\mathrm{A-D}}$. Pertinent reference measurements are shown in Figs. 2e-f, where the DC current $I_{\mathrm{A-D}}$ is flowing between contacts A and D, and contacts D and 1 are referenced to the ground. This resistance configuration shows the expected value of 0 at low currents, with a clear onset of dissipation observed when the measurement current $I_{\mathrm{A-D}}$ exceeds 620 nA, consistent with the Hall resistance measurement from Fig. 2b. Remarkably, in Fig. 2g-h, under the balanced scheme (with an additional current of the same magnitude $I_{\mathrm{4-1}}$ flowing into the inner perimeter), we observe that the longitudinal resistance signal remains below some 0.00002 $h/e^2$ (limited by the measurement resolution) up to measurement current of 3 $\mu$A.

\begin{figure}
\includegraphics[width=3in]{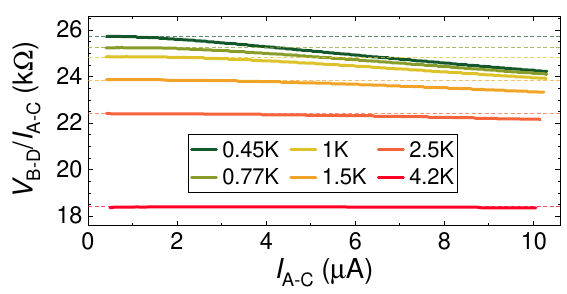}
\caption{
\textbf{Electrochemical potential balancing at various temperatures.} Hall resistance $V_{\mathrm{B-D}}$/$I_{\mathrm{A-C}}$ measured in the balanced scheme (Fig. 2c) at various temperatures, as a function of  $I_{\mathrm{A-C}}$. The colored horizontal dashed lines are guides to the eye, indicating the resistance at each temperature in the low current limit. Measurements are at zero external magnetic field and with 4V applied to the gate.
}%
\label{fig:Fig3}%
\end{figure}

To further support our interpretation that the slow drift away from the quantized value observed in Fig. 2d is caused by current induced heating, we consider measurements done at various higher temperatures on the same device, as shown in Fig. 3. It is well established for Cr/V-doped (Bi,Sb)$_{2}$Te$_3$ \cite{Bestwick2015,Fox2018,Fijalkowski2021b,Lippertz2022} that as the temperature is increased above some 100 mK, spurious conductance from the bulk of the material increases, the edge states become partially electrically shorted through the bulk, and the observed Hall resistance decreases. The data nevertheless shows that, as sample temperature is increased, the change in Hall resistance as a function of applied bias is progressively suppressed, as one would expect when the effects of current induced heating diminish. 

We note that the current at which we observe a breakdown without balancing in Figs. 2b and 2f (some 620 nA) corresponds to $\sim$16 mV of voltage difference between the perimeters $V_{\mathrm{B-2}}$. We find a clear breakdown of Hall resistance quantization whenever this voltage drop exceeds $\pm$16 mV, regardless of the measurement current $I_{\mathrm{A-C}}$ magnitude. In the multi-terminal Corbino geometry, this can be directly tested by connecting two independent voltage sources to the sample, one into each perimeter, and separately controlling the $I_{\mathrm{A-C}}$ and $I_{\mathrm{3-1}}$ currents. This is analyzed and discussed in detail in the supplementary material (Fig. S1), and further confirms that the electrochemical potential difference between the perimeters is the key parameter driving the QAHE quantization breakdown.

\section[Conclusions]{Conclusions}

As stated above, one can argue that there are two main obstacles to the QAHE entering mainstream metrology. The need for the device to support currents that are compatible with the high precision current comparators, and the need for the devices to work at liquid helium temperatures. The present work does not address the later issue, and only looks at the former.

Using a measurement scheme that drives the measurement current through each of the inner and outer edge of a multi-terminal Corbino ring, we demonstrate that balancing of the electrochemical potential between the two edges eliminates the electric field induced breakdown of the quantization. This is the breakdown that has thus far been responsible for limiting the current in QAHE based Hall bar devices.

By turning off this primary current-induced breakdown mechanism, we significantly increase the allowed measurement current before a secondary current-induced effect, that of Joule heating, becomes relevant. Importantly, the effect of Joule heating will be naturally suppressed when a measurement is performed in a dilution refrigerator with more cooling capacity. Indeed the inability to extract heat from the sample quickly enough is the likely reason for the observation of Joule heating in our current experimental setup. Therefore, while the experiment in a low cooling capacity system is appropriate as a proof of concept, we expect the maximum current the edge states can stably support to further increase with larger cooling power. We also note that the effects of Joule heating will be further mitigated by any progress that is made towards increasing the operating temperature, either in Cr/V-doped (Bi,Sb)$_{2}$Te$_3$, or other material systems. This allows us to conclude that the proposed measurement scheme fully takes care of one of the two limitations to the implementation of a QAHE based resistance standard. 

While the balancing wiring scheme presented here is easy to implement with standard measurement equipment, no instrument with metrologically relevant precision is currently configured for it. As such, a final assessment of the value of this method will require modification of the CCC systems.

Lastly, we note that the balancing method is not limited to QAHE based devices, and can also be used to increase the maximum current allowed in traditional quantum Hall based resistance devices, for applications where higher currents are useful. 

\section[Acknowledgents]{Acknowledgents}
The authors thank the team of H. Scherer (M. Kruskopf, M. G\"otz, D. Patel, and E. Pesel) from Physikalisch-Technische Bundesanstalt (PTB) Braunschweig for useful discussions about the metrological relevance of this work, and gratefully acknowledge the financial support of the Free State of Bavaria (the Institute for Topological Insulators), Deutsche Forschungsgemeinschaft (SFB 1170, 258499086), W\"urzburg-Dresden Cluster of Excellence on Complexity and Topology in Quantum Matter (EXC 2147, 39085490), and the European Commission under the H2020 FETPROACT Grant TOCHA (824140).

\section[Author contributions]{Author contributions}
K. M. F. together with C. G. developed the electrochemical potential balancing method. K. M. F. designed and patterned the measured devices, and performed the transport experiments. N. L. grew the MBE layer. M. K., S. S. and K. B. contributed to MBE material development. C. G., and L. W. M. supervised the work. All authors contributed to the analysis and interpretation of the results, and the writing of the manuscript.

\section[References]{References}


%

\clearpage
\setcounter{figure}{0} 
\renewcommand{\figurename}{Figure S}

\onecolumngrid

\large
\centerline{\textbf{Balanced Quantum Hall Resistor}}
\centerline{\textbf{- Supplementary Material}}
\normalsize
\bigskip
\centerline{K. M. Fijalkowski, N. Liu, M. Klement, S. Schreyeck, K. Brunner, C. Gould and L. W. Molenkamp}
\centerline{$^1$\textit{Faculty for Physics and Astronomy (EP3), Universit\"at W\"urzburg, Am Hubland, D-97074, W\"urzburg, Germany}}
\centerline{$^2$\textit{Institute for Topological Insulators, Am Hubland, D-97074, W\"urzburg, Germany}}
\bigskip

\setcounter{figure}{0} 
\renewcommand{\figurename}{Fig. S}

\section[Independent variation of the measurement current and the inter-edge electrochemical potential difference.]{Independent variation of the measurement current and the inter-edge electrochemical potential difference.}

Further insight into the mechanism driving the breakdown observed in Figs. 2b and 2f of the main text is obtained from independently biasing the outer and inner perimeter of the device. Fig. S1a shows the appropriate corresponding circuit diagram, where two independent DC voltage sources control the bias voltage, and therefore the current flowing into each perimeter. Contacts 1 and C are referenced to the ground, and therefore the electrochemical potential in each perimeter, at the position of contacts 4 and D is also determined by the circuit ground. The electrochemical potential at contacts A is carried by the outer perimeter edge state to contact B, whereas that of contact 3 is carried to 2.  The current flowing through each perimeter can be adjusted independently by tuning the voltage sources.

\medskip

The color scale in Fig. S1b indicates the four-terminal Hall resistance ($V_{\mathrm{B-D}}$/$I_{\mathrm{A-C}}$) measured for the outer perimeter. It is plotted as a function of the DC current through the outer edge $I_{\mathrm{A-C}}$, as well as the DC current through the inner edge $I_{\mathrm{3-1}}$. For better visibility of the colour scale, measurements where deviation from quantization exceeds 4 k$\Omega$ are grayed out in the figure. The main observation is immediately recognizable as the diagonal pattern in the figure, showing that the ideal quantized value (yellow color) is observed when the currents are comparable, whereas it quickly breaks down when the two currents differ significantly.

\medskip

To better highlight the role of the electric field, the same data can be plotted (Fig. S1c) as a function of the current through the outer edge $I_{\mathrm{A-C}}$, and the measured voltage $V_{\mathrm{B-2}}$ between the two perimeters. Shown in this way, it is clear that the voltage drop between perimeters is the key parameter behind the quantization breakdown. Whenever this voltage exceeds about $\pm$16 mV (marked with green vertical lines in the Figure), the value of $V_{\mathrm{B-D}}$/$I_{\mathrm{A-C}}$ quickly deviates from quantization.

\medskip

Fig. S1d-f shows an analogous analysis as Fig. S1a-c, but  for a four-terminal longitudinal resistance $V_{\mathrm{B-C}}$/$I_{\mathrm{A-D}}$ configuration. The results are consistent with that of the Hall signal: A clear diagonal pattern indicating the conditions where zero resistance is measured is observed in Fig. S1e, which persists for inter-edge voltages of up to $\pm$16 mV (Fig. S1f).

\medskip

As a consistency check, Fig S1g-h show the circuit and the measurement of a direct two terminal conductance between an inner and an outer edge contact.  For applied biases below some $\pm$16 mV, the bulk is insulating, and no current can flow, whereas conductance turns on quickly once this threshold value is reached. 

\begin{figure}[]
\includegraphics[width=\columnwidth]{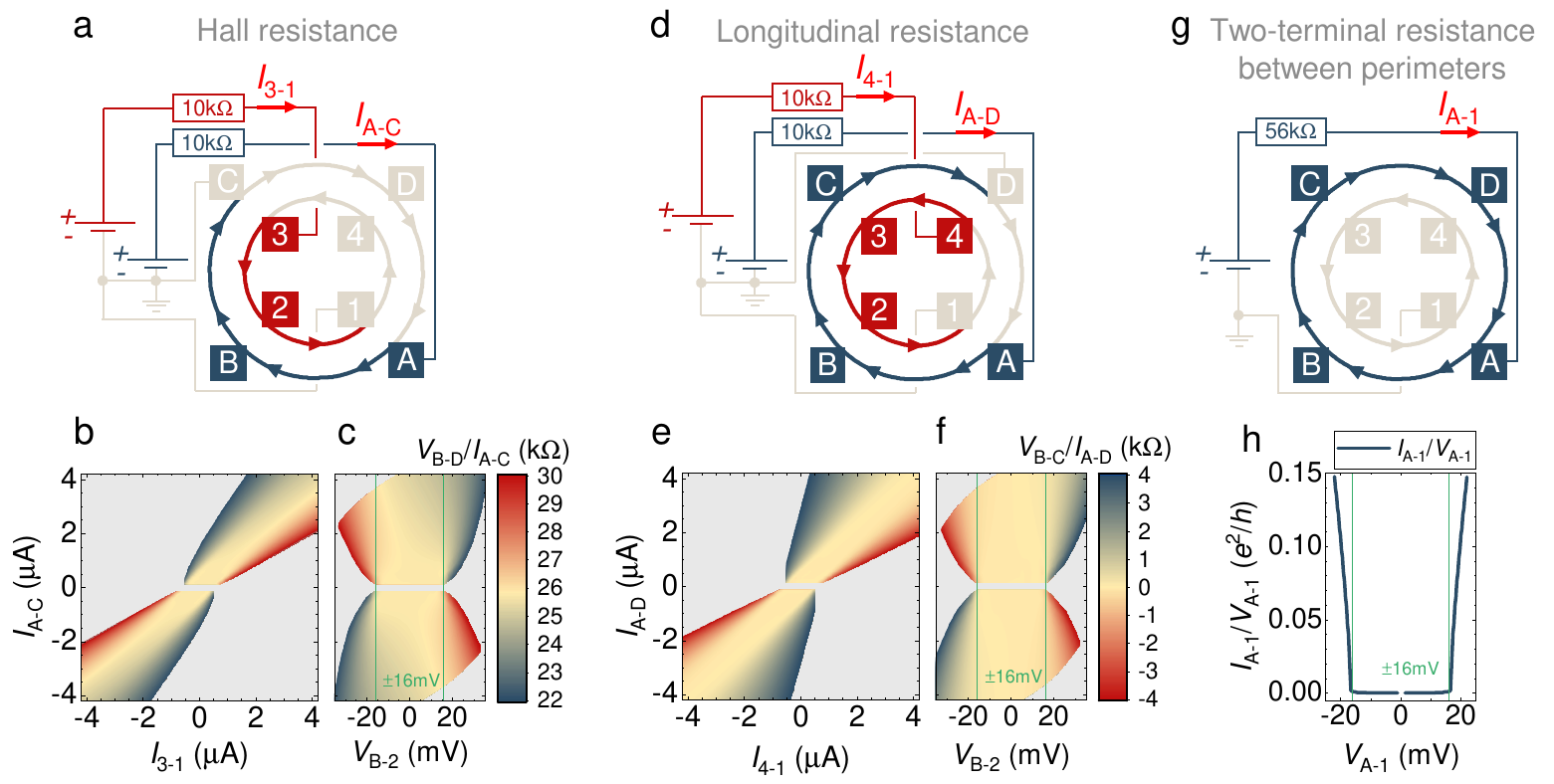}
\caption{
\textbf{Independent variation of the measurement current and the inter-edge electrochemical potential difference.} a) A circuit diagram for a measurement where two separate voltage sources are connected to the sample, one into each perimeter. This allows for an independent adjustment of the DC currents $I_{\mathrm{A-C}}$ and $I_{\mathrm{3-1}}$. The dark blue color represents the high electrochemical potential part of the circuit resulting from the bias voltage applied to the outer perimeter, whereas the dark red color represents the high potential part of the circuit resulting from the bias voltage applied to the inner perimeter. The light grey color represents the low electrochemical potential defined by circuit ground. b) Color plot showing how the four-terminal resistance $V_{\mathrm{B-D}}$/$I_{\mathrm{A-C}}$ changes with current in each perimeter. c) Color plot showing how $V_{\mathrm{B-D}}$/$I_{\mathrm{A-C}}$ changes with the voltage $V_{\mathrm{B-2}}$ and the measurement current $I_{\mathrm{A-C}}$. The green lines indicate the breakdown voltage of $\pm$16 mV. d-f) Equivalent analysis for a four-terminal (longitudinal) resistance $V_{\mathrm{B-C}}$/$I_{\mathrm{A-D}}$, with DC currents $I_{\mathrm{A-D}}$ and $I_{\mathrm{4-1}}$ flowing into the outer and inner perimeters respectively. g) A circuit diagram schematic for a two-terminal measurement where a bias voltage is applied directly between the two perimeters. h) A corresponding two-terminal conductance $I_{\mathrm{A-1}}$/$V_{\mathrm{A-1}}$ measurement plotted as a function of the applied bias voltage $V_{\mathrm{A-1}}$. Green lines show the breakdown voltage of $\pm$16 mV, consistent with the analysis in (c) and (f). All presented data was collected at zero external magnetic field, a temperature of 80 mK, and a gate voltage of 5.5V tuning the sample into the quantized plateau.
}
\label{fig:FigS1}
\end{figure}

\end{document}